\documentclass[aps,preprintnumbers,nofootinbib,superscriptaddress,11pt]{revtex4}
\usepackage[hypertex]{hyperref}
\usepackage[dvipdfmx]{graphicx} 
\usepackage{amsmath,amssymb,amsfonts,bm,cancel} 

            \def\l{\lambda}  \def\m{\mu} \def\n{\nu}        \def\t{\tau}       

\def\dg{\dagger}  \def\nn{\nonumber}

\newcommand{\lsp}{ \left ( } \newcommand{\rsp}{ \right ) }     \newcommand{\getsto}{\leftrightarrow}

\def\abs#1{\left| #1\right|}

\newcommand{\row}[2]{ \begin{pmatrix}  #1 & #2   \end{pmatrix}  }
\newcommand{\column}[2]{ \begin{pmatrix}  #1 \\ #2 \\  \end{pmatrix} }

\newcommand{\diag}[2]{ \begin{pmatrix}  #1 & 0 \\ 0 & #2 \\   \end{pmatrix}  }

\begin{document}


\title{\Large Chiral perturbative solution for the type-I seesaw mechanism \\ in next-to-leading order}

\preprint{STUPP-23-269}
\author{Masaki J. S. Yang}
\email{mjsyang@mail.saitama-u.ac.jp}
\affiliation{Department of Physics, Saitama University, 
Shimo-okubo, Sakura-ku, Saitama, 338-8570, Japan}



\begin{abstract} 

In this letter, we perform chiral perturbative diagonalization of the type-I seesaw mechanism by hierarchical singular values $\l_{i}$ of the Dirac mass matrix $m_{D}$ up to the next-to-leading order (NLO). 
Since the mass matrix of right-handed neutrinos $M_{R}$ has parity symmetries under $\l_{i} \getsto - \l_{i}$, 
the singular values $M_{i}$ and mixing angles of diagonalization of $M_{R}$ are written by only even and odd orders of $\l_{i}$ respectively.

We confirm this fact by specific perturbative expansions of the third- and the fourth-order by $\l_{i}$. 
As a result,  as long as the chiral perturbation theory is valid, 
the NLO contributions are generally suppressed by $O(\l_{i}^{2} / \l_{j}^{2}) \lesssim 1\%$ compared to the leading-order expressions that have sufficient accuracies.

\end{abstract} 

\maketitle

\section{Introduction}

The chiral perturbation theory \cite{Gasser:1984gg} is a highly useful approach for relating the masses of mesons and quarks. 
In the type-I seesaw mechanism \cite{Minkowski:1977sc, GellMann:1980v, Yanagida:1979as, Mohapatra:1979ia}, 
the lepton number symmetry which is restored in the massless limit of a neutrino plays the role of chiral symmetry 
 \cite{Wyler:1982dd, Petcov:1984nz,  Branco:1988ex, Kersten:2007vk, Adhikari:2010yt}. 
Recently, by applying chiral perturbative treatment to the seesaw mechanism, we can calculate the masses of right-handed neutrinos $M_{i}$ and their mixings \cite{Yang:2022bex, Yang:2023ixv, Yang:2023aqh, Yang:2023cso}. While the perturbation theory for quarks has been established up to higher-order contributions, it is not known how significant such higher-order contributions are for the seesaw mechanism.

Therefore, in this paper, we proceed with the chiral perturbative treatment for the type-I seesaw mechanism up to 
next-to-leading order (NLO) to evaluate errors of the leading order expressions. 
Since right-handed neutrinos $\n_{R i}$ play important roles in grand unified theories and leptogenesis, 
these evaluations will be useful in phenomenological analyses.

%
%

\section{Chiral perturbative diagonalization of $M_{R}$}

In this section, we review the chiral perturbative diagonalization of the mass matrix $M_{R}$ of the right-handed neutrinos $\n_{R i}$ \cite{Yang:2022bex, Yang:2023ixv, Yang:2023aqh, Yang:2023cso}.
The Dirac mass matrix $m_D$ and the mass matrix of light neutrinos $m_{\n} = U_{\rm MNS} m_{\n}^{\rm diag} U_{\rm MNS}^{T}$ are defined in a basis where the Yukawa matrix of charged leptons $Y_e$ and $M_R$ are diagonal.
Here, $U_{\rm MNS}$ is the MNS matrix.

We define the singular value decomposition (SVD) as $m_{D} = V m_{D}^{\rm diag} U^{\dg}$ with its singular values $\l_{i}$ and 
proper unitary matrices $V, U$. 
In the diagonal basis of $m_D$, the type-I seesaw mechanism is expressed as 
\begin{align}
 m_{D}^{\rm diag} V^{T} m_{\n}^{-1} V m_{D}^{\rm diag } 
& = U^{T} M_{R}^{\rm diag } U  \, . 
\end{align}
In this basis, by defining the mass matrix $V^{T} m_{\n}^{-1} V \equiv m^{-1}$ (and $V^{\dg} m_{\n} V^{*} = m $), 
the matrix elements of $M_R$ are written by $(M_{R})_{ij} = \l_{i } (m^{-1})_{ij} \l_{j}$. 
If the singular values of $m_{D}$ are hierarchical ($\l_{1} \ll \l_{2} \ll \l_{3}$) and 
$m$ has an inverse matrix (that is, the lightest mass is not zero $m_{1 \, \rm or \, 3 } \neq 0$), 
the diagonalization of $M_R$ can be performed chiral perturbatively. 
First, after integrating out the heaviest third-generation $\n_{R3}$, 
the mass matrix of the remaining two lighter generations $M_{R0}$ is expressed as 
\begin{align}
M_{R0} 
 & = 
\begin{pmatrix}
\l_{1}^{2} (m^{-1})_{11} & \l_{1} \l_{2} (m^{-1})_{12} \\
 \l_{1} \l_{2}  (m^{-1})_{12}  & \l_{2}^{2} (m^{-1})_{22} 
\end{pmatrix} 
 - {1\over \l_{3}^{2} (m^{-1})_{33}} \l_{3}^{2}
\column{\l_{1}  (m^{-1})_{13} }{\l_{2} (m^{-1})_{23} } \otimes
\row{\l_{1}  (m^{-1})_{13} }{\l_{2}  (m^{-1})_{23} } \nn \\
 & = {1\over m_{11} m_{22} - m_{12}^{2}}
\diag{\l_{1}}{\l_{2}}
\begin{pmatrix}
m_{22} & - m_{12} \\ - m_{12} & m_{11}
\end{pmatrix} 
\diag{\l_{1}}{\l_{2}} . 
\label{MR0}
\end{align}
It essentially reduces to a simpler seesaw mechanism for the remaining two generations.

By integrating out the second heaviest neutrino $\n_{R2}$, 
$m^{-1}$ becomes a "$1 \times 1$ inverse matrix" $1 / m_{11}$, and the diagonalized elements $M_{i}$ of $M_{R}$ are given by 
\begin{align}
(M_{1} \, , \, M_{2} \, , \, M_{3}) \simeq 
 ( {\l_{1}^{2} \over m_{11} } \, ,  \,   {\l_{2}^{2} \, m_{11} \over m_{11} m_{22} - m_{12}^{2} } \, , \,  \l_{3}^{2} (m^{-1})_{33} ) \, . 
\label{Mi}
\end{align}
The singular values of $M_{R}$ are the absolute values of Eq.~(\ref{Mi}).

The diagonalization matrix of $M_{R}$ associated with the integration is 
\begin{align}
U &\simeq 
\begin{pmatrix}
 1 &   {\l_{1} \over \l_{2}} {m_{12}^{*} \over m_{11}^{*}} & 0 \\
- {\l_{1} \over \l_{2}}   {m_{12} \over m_{11} } & 1 & 0 \\
 0 & 0 & 1 \\
\end{pmatrix} 
\begin{pmatrix}
 1 & 0 & - {\l_{1} \over \l_{3}} {(m^{-1})^{*}_{13} \over (m^{-1})^{*}_{33} } \\
 0 & 1 & - {\l_{2} \over \l_{3}}  { (m^{-1})^{*}_{23} \over (m^{-1})^{*}_{33} } \\
  {\l_{1} \over \l_{3}}  { (m^{-1})_{13} \over (m^{-1})_{33} } & {\l_{2} \over \l_{3}} { (m^{-1})_{23} \over (m^{-1})_{33} } & 1 \\
\end{pmatrix}  \label{Eq6} \\ 
& \simeq 
\begin{pmatrix}
 1 &  {\l_{1} \over \l_{2}} {m_{12}^{*} \over m_{11}^{*}} &  {\l_{1} \over \l_{3}} {m_{13}^{*} \over m_{11}^{*}}  \\ 
- {\l_{1} \over \l_{2}}   {m_{12} \over m_{11} } & 1 &  - {\l_{2} \over \l_{3}}  { (m^{-1})^{*}_{23} \over (m^{-1})^{*}_{33} }  \\
  {\l_{1} \over \l_{3}}  { (m^{-1})_{13} \over (m^{-1})_{33} } & {\l_{2} \over \l_{3}} { (m^{-1})_{23} \over (m^{-1})_{33} } & 1 \\
\end{pmatrix} \, . 
\end{align}
This procedure is equivalent to the second-order perturbation theory of SVD \cite{Yang:2022bex} 
and perturbatively "solves" all six constraints of the type-I seesaw mechanism \cite{Yang:2023aqh}. 
In other words, one can determine $U$ and $M_{i}$ from input parameters  $m_{\n}$, $m_{Di}$, and $V$ \cite{Davidson:2001zk}.

For chiral perturbative diagonalization to be a good description, 
$U$ must not have large mixings and the following conditions must be satisfied;
\begin{align}
{\l_{1} \over \l_{2}}  \abs{m_{12} \over m_{11} } \, , ~   {\l_{1} \over \l_{3}} \abs { (m^{-1})_{13} \over (m^{-1})_{33} } \, , ~  {\l_{2} \over \l_{3}} \abs{ (m^{-1})_{23} \over (m^{-1})_{33} } \, \lesssim \,  0.1 \, . 
\end{align}
As long as the parameter $m_{11}$ and the lightest mass $m_{1 \rm \, or \, 3}$ are not too small, 
this perturbative treatment is effective over a wide parameter range. 
Errors of perturbations are clearly of the second or lower order of $\l_{i}$. 
The next section shows that the errors of mixing angles are of the third order or less.

\section{Chiral perturbation theory of $M_{R}$ in next-to-leading order}

Here, to evaluate the NLO contribution, 
we perform the perturbative treatment partially in the third- and fourth-order.

\subsection{Overview: Parity of $\l_{i}$}

Before a detailed calculation, we overview 
that parities associated with the singular values $\l_{i}$ strongly constrain the SVD of $M_{R}$.
First, it is apparent that the singular values of $M_{R}$ have only even orders of $\l_{i}$. 
For simplicity, by considering only two generations as in Eq.~(\ref{MR0}), 
the Hermitian matrix $M_{R0} M_{R0}^{\dg}$ is
\begin{align}
M_{R0} M_{R0}^{\dg} = {1\over |\det m_{0}|^{2}}
\begin{pmatrix}
\l_{1}^4 |m_{22}|^{2} + \l_{1}^{2} \l_{2}^2 |m_{12}|^{2} & 
- \l_{1}^3 \l_{2} m_{22} m_{12}^* - \l_{1} \l_{2}^3 m_{12} m_{11}^* \\
- \l_{1}^3 \l_{2} m_{12} m_{22}^* - \l_{1} \l_{2}^3 m_{11} m_{12}^* & 
\l_{1}^2 \l_{2}^{2} |m_{12}|^{2} + \l_{2}^4 |m_{11}|^{2} \\
\end{pmatrix} , 
\end{align}
where $\det m_{0} \equiv m_{11} m_{22} - m_{12}^{2}$. 
Its diagonalization can be solved exactly \cite{Endoh:2002wm}. 
Since the characteristic equation involves only even orders of $\l_{i}$, 
the eigenvalues are written by functions of $\l_{i}^{2}$. 
This feature also holds for the case of three generations, 
because the matrix elements are written by $(M_{R}^{\dg} M_{R})_{ik} = \sum_{j} \l_{i} (m^{-1})^{\dg}_{ij} \l_{j}^{2} (m^{-1})_{jk} \l_{k}$. 

Next, the unitary matrix $U$ associated with the SVD is constrained by parities of $\l_{i}$.
We consider parity transformations $p_{i}:\l_{i} \to - \l_{i}$ that change the sign of $\l_{1,2,3}$ 
and define $P \equiv {\rm diag} (p_{1} \, , \, p_{2} \, , \, p_{3})$.
Since only non-diagonal elements have odd orders in $\l_{i}$, the mass matrix for three generations $M_{R}$ and diagonalized one $M_{R}^{\rm diag}$ 
have the symmetry under these transformations;   
\begin{align}
P M_{R} P =  M_{R} \, , ~~ P M_{R}^{\rm diag} P = M_{R}^{\rm diag} \, . 
\end{align}
By using $P^{2} = 1$, the following holds from the SVD.
\begin{align}
 M_{R}^{\rm diag} = U^{T} M_{R} U  =  P U^{T} P M_{R} P U P  \, .
\end{align}
Since these symmetries hold for any $\l_{i}$, $P U P = U$ is required, and the unitary matrix $U$ also has these symmetries. 
That is, the diagonal elements have even orders of $\l_{i}$ and 
the non-diagonal $ij$ elements have only odd orders of the corresponding $\l_{i} \l_{j}$.

From this, the perturbative expansion of $U$ becomes 
\begin{align}
U \simeq 
\begin{pmatrix}
1 + \dfrac{\l_{i}^{2}}{\l_{j}^{2}} & \dfrac{\l_{1}}{\l_{2}} + \dfrac{\l_{1} \l_{2} }{\l_{3}^{2}} + \dfrac{\l_{1}^{3}}{\l_{2}^{3}} & \dfrac{\l_{1}}{\l_{3}} + \dfrac{ \l_{1} \l_{2}^{2}}{\l_{3}^{3}}   \\
\dfrac{\l_{1}}{\l_{2}} + \dfrac{\l_{1} \l_{2} }{\l_{3}^{2}} + \dfrac{\l_{1}^{3}}{\l_{2}^{3}} & 1 +  \dfrac{\l_{i}^{2}}{\l_{j}^{2}} & \dfrac{\l_{2}}{\l_{3}} + \dfrac{\l_{2}^{3}}{\l_{3}^{3}} \\
\dfrac{\l_{1}}{\l_{3}} + \dfrac{ \l_{1} \l_{2}^{2}}{\l_{3}^{3}}  & \dfrac{\l_{2}}{\l_{3}} + \dfrac{\l_{2}^{3}}{\l_{3}^{3}}  & 1 +  \dfrac{\l_{i}^{2}}{\l_{j}^{2}} \\
\end{pmatrix} ,
\end{align}
where each term represents the order of the perturbation and $O(1)$ coefficients are ignored.
In the 2-3 and 3-2 elements, there is also a term of  $\l_{1}^{2} \l_{2} / \l_{3}^{3}$. 
However, this term (that does not emerge in the two-generational case) is not treated as NLO 
because this is suppressed by $\l_{1}^{2} / \l_{2}^{2}$ compared to the term $\l_{2}^{3} / \l_{3}^{3}$.

On the other hand, the 1-2 element has two NLO contributions
\begin{align}
{\l_{1} \over \l_{2}} {\l_{2}^{2} \over \l_{3}^{2}} = \frac{\l_{1} \l_{2} }{\l_{3}^{2}} \, , ~~~
{\l_{1} \over \l_{2}} {\l_{1}^{2} \over \l_{2}^{2}} =  \frac{\l_{1}^{3}}{\l_{2}^{3}} \, .
\end{align}
%
%
The values of $\l_{i}$ correspond to fermion masses of the Standard Model (SM) by some grand unified relations.  
In the SM, the ratios 
$\l_{1} / \l_{2}$ and $\l_{2} / \l_{3}$ are approximately \cite{Xing:2007fb}
\begin{align}
{m_{u} \over m_{c} } \simeq {1 \over 500} \, ,  ~&~ {m_{d} \over m_{s} } \simeq  {1 \over 20} \, ,  ~~ {m_{e} \over m_{\m} } \simeq {1 \over 200 } \, ,  \\
{m_{c} \over m_{t} } \simeq {1 \over 300} \, ,  ~&~ {m_{s} \over m_{b} } \simeq  {1 \over 50} \, ,  ~~ {m_{\m} \over m_{\t} } \simeq {1 \over 16 } \, .
\end{align}
Since we cannot determine which term is dominant, both terms are treated as NLOs in this paper.
Nevertheless, the values of these second-order perturbations are 
\begin{align}
{m_{u}  m_{c} \over m_{t}^{2} } \simeq { 1\over  5 \times 10^{7}} \, , ~~~
{m_{d} m_{s} \over m_{b}^{2} } \simeq  {1 \over 50000} \, , ~~~ 
{m_{e} m_{\m} \over m_{\t}^{2} } \simeq {1 \over 50000 } \, ,
\end{align}
and any combination is less than 0.01\%.

\subsection{Diagonalization of two generations}

%
Hereafter, we will consider partial third- and fourth-order perturbation theory 
that evaluates NLO contributions.
Let us consider the simpler two-generation case when the heaviest $\n_{R3}$ is integrated out.
The diagonalization matrix in the second-order of $\l_{1}$ is defined by 
\begin{align}
U_{0} = 
\begin{pmatrix}
c & s^{*}  \\
- s & c \\
\end{pmatrix} ,  ~~ 
s = - \frac{\l_{1} m_{12}}{\l_{2} m_{11}} 
 \, ,
~~
c = 1- {1\over 2}\frac{ \l_{1}^2 |m_{12}|^{2}}{\l_{2}^2 |m_{11}|^{2} } \, .
\end{align}
This $U_{0}$ transforms the mass matrix~(\ref{MR0}) as 
\begin{align}
U_{0}^{T} M_{R0} U_{0} = 
\begin{pmatrix}
\dfrac{ \l_{1}^2  }{m_{11}} &  \dfrac{\l_{1}^3}{\l_{2}} B  \\[10pt]
\dfrac{\l_{1}^3}{\l_{2}} B  & \l_{2}^2 \dfrac{m_{11} }{ \det m_{0}} \lsp 1 + \dfrac{\l_{1}^2}{\l_{2}^{2}} \dfrac{| m_{12}|^2}{ |m_{11}|^2} \rsp \\
\end{pmatrix} + O \lsp \dfrac{\l_{1}^{4}}{ \l_{2}^{2}} \rsp \, , 
\end{align}
where 
\begin{align}
B = - \frac{ m_{12}^{*} }{ |m_{11}|^2 } + {1\over 2} \frac{ m_{12}}{ \det m_{0}} \frac{|m_{12}|^2}{|m_{11}|^2}
\, . 
\end{align}
From the parity of $\l_{1}$, the non-diagonal element has no second-order perturbation.

To find the lighter singular value $M_{1}$ of $M_{R0}$ requires a fourth-order evaluation of $\l_{1}$.
However, this $M_{1}$ is easily obtained from the relation between determinants $M_{1} M_{2} = \l_{1}^{2} \l_{2}^{2} / \det m_{0}$; 
\begin{align}
M_{1} \simeq  \frac{\l_{1}^2  }{m_{11}} \lsp 1 - \dfrac{\l_{1}^{2} }{ \l_{2}^{2}} \dfrac{| m_{12}|^2 }{| m_{11}|^2 }  \rsp \, , ~~
M_{2} \simeq \l_{2}^2 \dfrac{m_{11} }{ m_{11} m_{22} - m_{12}^{2}} \lsp 1 + \dfrac{\l_{1}^2}{\l_{2}^{2}} \dfrac{| m_{12}|^2}{ |m_{11}|^2} \rsp \, . 
\end{align}
Thus, the NLO corrections to $M_{i}$ are suppressed by $\l_{1}^{2} m_{12} / \l_{2}^{2} m_{11}$.

Also, the third-order correction to $U_{0}$ is
\begin{align}
U_{1} = 
\begin{pmatrix}
1 &  \dfrac{\l_{1}^3  }{\l_{2}^{3}} \dfrac{B^{*} \det m_{0}}{m_{11}^{*}} \\
-  \dfrac{\l_{1}^3 }{\l_{2}^{3} } \dfrac{B \det m_{0}}{m_{11}} & 1 
\end{pmatrix} , ~~~
%
\end{align}
and is approximately $\l_{1}^{3} m_{12}^{3} / \l_{2}^{3} m_{11}^{3}$.
These results coincide with a series expansion of the exact solution of SVD.

The perturbation theory breaks down in the limit of $m_{11} \to 0$, 
because the diagonalization of the lighter generation $M_{R0}$~(\ref{MR0}) has a large mixing.
In this case, the relation between the masses $M_{1}^{-1} \propto m_{11} \propto M_{2}$ no longer holds,
and the diagonalization must be done more accurately.  This situation is discussed in Ref.~\cite{Yang:2023cso}.

\subsection{Diagonalization of three generations}

We perform a similar calculation for the case of three generations. 
The mass matrix of right-handed neutrinos $M_{R}$ is
\begin{align}
M_{R} 
= 
\begin{pmatrix}
\l_1^2 (m^{-1})_{11} & \l_1 \l_2 (m^{-1})_{12} & \l_1 \l_3 (m^{-1})_{13} \\
\l_1 \l_2 (m^{-1})_{12} & \l_2^{2} (m^{-1})_{22} & \l_2 \l_3 (m^{-1})_{23}  \\
\l_1 \l_3 (m^{-1})_{13}  & \l_2 \l_3 (m^{-1})_{23} & \l_3^2 (m^{-1})_{33} \\
\end{pmatrix} .
\end{align}
The diagonalization matrix in the second-order of $\l_{i}$ is
\begin{align}
U_{2} = U_{23} U_{13} U_{12} 
= 
\begin{pmatrix}
1 & 0 & 0 \\
0 & c_{23} & s_{23}^{*} \\
0 & -s_{23} & c_{23}
\end{pmatrix}
\begin{pmatrix}
c_{13} & 0 & s_{13}^{*} \\
0 & 1 & 0 \\
- s_{13} & 0 & c_{13}
\end{pmatrix}
\begin{pmatrix}
c_{12} & s_{12}^{*} & 0 \\
- s_{12} & c_{12} & 0 \\
0 & 0 & 1
\end{pmatrix}  , 
\end{align}
where $c_{ij} = 1 -{1\over 2} |s_{ij}|^{2}$ and
\begin{align}
s_{23} = {\l_{2} \over \l_{3}} {(m^{-1})_{23} \over (m^{-1})_{33}} \, , ~~
s_{13} = {\l_{1} \over \l_{3} }{(m^{-1})_{13} \over (m^{-1})_{33}} \, , ~~ 
s_{12} = - {\l_{1} \over \l_{2} } {m_{12} \over m_{11}}
 \, .
\end{align}
This $U_{2} \simeq U^{\dg}$ corresponds to the integration out described in Eq.~(\ref{Eq6}). 

By a basis transformation of $U_{2}$, all off-diagonal elements of $M_{R}$ become third-order perturbations, 
\begin{align}
U_{2}^{T} M_{R} U_{2} \simeq  
\begin{pmatrix}
\dfrac{\l_{1}^{2} }{m_{11}} & \dfrac{\l_{1}^{3}}{\l_{2}} B_{12} + \dfrac{\l_{1} \l_{2}^{3}}{\l_{3}^{2}} C_{12} &  \dfrac{\l_{1} \l_{2}^{2}}{\l_{3}} B_{13} \\[10pt]
 \dfrac{\l_{1}^{3}}{\l_{2}} B_{12} + \dfrac{\l_{1} \l_{2}^{3}}{\l_{3}^{2}} C_{12} & \dfrac{\l_{2}^{2} \, m_{11}}{\det m_{0}} (1 + |s_{12}|^{2}) &  \dfrac{\l_{2}^{3}}{\l_{3}} B_{23} \\[10pt]
\dfrac{\l_{1} \l_{2}^{2}}{\l_{3}} B_{13} &  \dfrac{\l_{2}^{3}}{\l_{3}} B_{23} & 
{\l^{3} (m^{-1})_{33}} ( 1 + |s_{23}|^{2} + |s_{13}|^{2} )
\end{pmatrix} , 
\end{align}
where $B_{ij}$ and $C_{12}$ are 
\begin{align}
B_{23} & =  \frac{ (m^{-1})_{23}^{*} }{  |(m^{-1})_{33}|^2} \lsp  \frac{ m_{11}  }{ \det m } - {1\over 2} { (m^{-1})_{23}^{2} } \rsp \, ,   \\
B_{13} & = \frac{ (m^{-1})_{23}^{*} }{  |(m^{-1})_{33}|^2} \lsp - \frac{ m_{12}  }{ \det m } - {1\over 2} { (m^{-1})_{13} (m^{-1})_{23} } \rsp  \, ,  \\
B_{12} &= B =  \frac{ m_{12}^{*} }{ |m_{11}|^2 } \lsp -1 + {1\over 2} \frac{ m_{12}^{2}}{m_{11} m_{22} - m_{12}^{2}} \rsp \, , \\
C_{12} & = { (m^{-1})_{23}^*\over 2 \det m | (m^{-1})_{33}|^2} \lsp
 m_{13} - m_{11} \frac{(m^{-1})_{13} }{ (m^{-1})_{33}}  \rsp
  \, . 
\end{align}

In denominators, some of parameters have possibilities of divergence. 
Since the term $(m^{-1})_{33}$ approximately satisfies $1/m_{33} \sim 2/m_{3}$, 
it is safe in the normal hierarchy (NH). 
The limit $\det m \to 0$ (that means $m_{1 \, \rm or \, 3} \to 0$) contains the case of the inverted hierarchy (IH). 
If the limit corresponds to $M_{3} \to \infty$, this case can be treated perturbatively by removing the divergent part \cite{Yang:2023cso}.
Except for this situation, the chiral perturbation theory is valid as long as $m_{11}$ is not close to zero.

Since $B_{ij}$ and $C_{12}$ are about $1/m_{ij}$, their corrections on $U_{2}$ are about $B_{ij} / M_{j}$ and $C_{12} / M_{2}$, 
which is suppressed by $\l_{i}^{2} / \l_{j}^{2}$ compared to LO contributions of mixing angles.
Furthermore, effects  on $M_{i}$ from these third-order perturbations are the sixth order and 
we can ignore all contribution except on $M_{1}$ (there is a term of $B_{12} C_{12} \l_{1}^{4} / \l_{3}^{2}$ from the cross term).

By comparing the determinants as in the two-generational case, we obtain diagonal elements of $M_{R}$; 
\begin{align}
M_{1} &\simeq \dfrac{\l_{1}^{2}}{m_{11}} \lsp 1 -  {\l_{1}^{2} \over \l_{2}^{2}} \abs{ m_{12} \over m_{11}}^{2} - {\l_{1}^{2} \over \l_{3}^{2}} \abs{(m^{-1})_{13} \over (m^{-1})_{33}}^{2} \rsp  \, , \\
M_{2} &\simeq \dfrac{\l_{2}^{2} \, m_{11}}{m_{11} m_{22} - m_{12}^{2}}
 \lsp 1 +  {\l_{1}^{2} \over \l_{2}^{2}} \abs{ m_{12} \over m_{11}}^{2} 
 - {\l_{2}^{2} \over \l_{3}^{2}} \abs{(m^{-1})_{23} \over (m^{-1})_{33}}^{2}  \rsp \, , \\
M_{3} &\simeq {\l_{3}^{2} (m^{-1})_{33}} 
 \lsp 1 + {\l_{1}^{2} \over \l_{3}^{2}} \abs{(m^{-1})_{13} \over (m^{-1})_{33}}^{2} + {\l_{2}^{2} \over \l_{3}^{2}} \abs{(m^{-1})_{23} \over (m^{-1})_{33}}^{2}\rsp \, . 
\end{align}
Therefore, the NLO corrections to $M_{i}$ are of $O(\l_{i}^{2} / \l_{j}^{2})$ 
(Actually, the term $\l_{1}^{2} / \l_{3}^{2}$ is very small and can be ignored). 
As a result, if the singular values $\l_{i}$ are about fermion masses of the SM, 
$\l_{i}^{2} / \l_{j}^{2} < 1 \%$ holds for any combination.
Errors of the chiral perturbation theory are smaller than that of neutrino oscillation experiments and the leading order expression is sufficiently accurate.

Other physical quantities in the seesaw mechanism, such as $m_{D}$ and the CP-violating matrix $m_{D}^{\dg} m_{D}$, are also sufficiently precise in the leading order evaluation. 
For example, the complex orthogonal matrix $R$ in Casas--Ibarra parameterization \cite{Casas:2001sr} is 
represented as 
$R_{i1} \simeq \pm \sqrt{m_{i} / m_{11} } (U_{\rm MNS}^{T} V^{*})_{i1} \, ,  
 R_{i2} \simeq \pm \sum_{j,k} \epsilon_{ijk} R_{j3} R_{k1} \, , 
 R_{i3} \simeq  \pm {(U_{\rm MNS}^{\dagger} V)_{i3} / \sqrt {m_{i} (m^{-1})_{33}} } $ 
in the leading order of $\l_{i}$ \cite{Yang:2023aqh}. 

From the parity of $\l_{i}$ for $R$, errors for this representation are also suppressed by $O(\l_{i}^{2} / \l_{j}^{2})$. 
 First, the definition $m_{D} = U_{\rm MNS} \sqrt{m_{\n}^{\rm diag}} R \sqrt{M_{R}^{\rm diag}}$ yields a relation
\begin{align}
R = \sqrt{m_{\n}^{\rm diag -1}} U_{\rm MNS}^{\dg} V m_{D}^{\rm diag} U^{\dg} \sqrt{M_{R}^{\rm diag -1}} \, . 
\end{align}
Since $M_{i}$ is proportional to $\l_{i}^{2}$ and has only even orders of $\l_{i}$, 
$\sqrt{M_{R}^{\rm diag}}$ (and its inverse matrix) has only odd orders of $\l_{i}$.
Therefore, the matrix $M_{D}^{\rm diag} U^{\dg} \sqrt{M_{R}^{\rm diag -1}}$ and $R$ has only even orders of $\l_{i}$, 
and errors are expected to be about $O(\l_{i}^{2} / \l_{j}^{2})$, less than 1 \%. 
 
Finally, the masses $M_{i}$ and mixings associated with approximate chiral symmetries are hardly renormalized \cite{tHooft:1979bh}. 
Parameters involving third generations ($m_{i3}$ and $\l_{3}$) may be changed by quantum corrections at most 10\% 
even when the Yukawa interactions of tau lepton $y_{\t}$ and $\l_{3}$ are about that of top quark \cite{Xing:2007fb}.

\section{summary}

In this letter, we perform chiral perturbative diagonalization of the type-I seesaw mechanism by hierarchical singular values $\l_{i}$ of the Dirac mass matrix $m_{D}$ up to the next-to-leading order (NLO). 
Since the mass matrix of right-handed neutrinos $M_{R}$ has parity symmetries under $\l_{i} \getsto - \l_{i}$, 
the singular values $M_{i}$ and mixing angles of diagonalization matrix $U$ of $M_{R}$ are written by only even and odd orders of $\l_{i}$ respectively.

We confirm this fact by specific perturbative expansion of the third- and the fourth-order by $\l_{i}$. 
As a result, 
as long as the chiral perturbation theory is valid ($m_{11} \neq 0$ and $m_{\rm 1 \, or \, 3} \neq 0$), 
the NLO contributions are generally suppressed by $O(\l_{i}^{2} / \l_{j}^{2}) \lesssim 1\%$ compared to the leading-order expressions that have sufficient accuracies.
The same is true for the complex orthogonal matrix $R$ and the Dirac mass matrix $m_{D}$ displayed in the chiral perturbation theory.

Right-handed neutrinos play important roles in grand unified theory and leptogenesis \cite{Fukugita:1986hr}, and the parameter $m_{11}$ appearing in the chiral perturbation theory is related to the effective mass of double beta decay $m_{ee}$. 
The sufficient accuracy of the chiral perturbation theory will be useful in phenomenological analysis and cosmology.


\end{document}